\documentclass{article}
\input{psfig.sty}
\newcommand{\eq}{\begin{equation}}
\newcommand{\en}{\end{equation}}
\newcommand{\eqa}{\begin{eqnarray}}
\newcommand{\ena}{\end{eqnarray}}
\begin{document}
\baselineskip 1cm

\title{Redundancy and synergy arising from  correlations in large ensembles}
\author{Michele Bezzi, Mathew E. Diamond and Alessandro Treves \\ 
SISSA - Programme in Neuroscience \\
via Beirut 4, 34014 Trieste, Italy}
\date{\today}
\maketitle

\begin{abstract}

Multielectrode arrays allow recording of 
the activity of many single neurons, from which correlations
can be calculated.
The functional roles of correlations can be revealed by the measures
of the information conveyed by neuronal activity;  a simple formula 
has been shown to discriminate the information transmitted by individual spikes from the
positive or negative contributions due to correlations
(Panzeri et al, Proc. Roy. Soc. B.,  {\bf 266}: 1001--1012 (1999)).
The formula  quantifies the
corrections to the single-unit instantaneous information rate which result 
from correlations in spike emission between pairs of neurons. 
Positive corrections imply synergy, while negative corrections 
indicate redundancy. Here, this analysis,  previously  applied 
to recordings from small ensembles, is developed further by considering 
a model of a large ensemble, in which correlations among the signal and 
noise components of neuronal firing are small in absolute value 
and entirely random in origin. Even such 
small random correlations are shown to lead to large possible synergy or 
redundancy, whenever the time window for extracting information 
from neuronal firing extends to the order of the mean interspike 
interval. In addition, a sample of recordings from rat barrel cortex
illustrates the mean time window at which such `corrections' dominate when correlations
are, as often in the real brain, neither random nor small. 
The presence of this kind of correlations for a large ensemble of cells
restricts further the time of validity of the expansion,
 unless what is decodable by the receiver 
 is also taken into account.
 
\end{abstract}
\section{Do correlations convey more information than do rates alone?}

Our intuition often brings us to regard
neurons as {\em independent} actors in the business of information processing.
We are then reminded of the potential for intricate mutual dependence in
their activity, stemming from common inputs and from interconnections,
and are finally brought to consider correlations as sources of much 
richer, although somewhat hidden, 
information about what a neural ensemble is really doing. Now that
the recording of multiple single units is common practice in many laboratories,
correlations in their activity can be measured and their
role in information processing can be elucidated case by case. 
Is the information conveyed by the activity of an ensemble of neurons 
determined solely by the number of spikes fired by each cell as could be
quantified also with non-simultaneous recordings~\cite{Rol+97}; or do 
correlations in the emission of action potentials also play a significant 
role?

Experimental evidence on the role of correlations in neural coding of
sensory events, or of internal states, has been largely confined to ensembles
of very few cells. Their contribution has been said to be positive, i.e.
the information contained in the ensemble response is greater than 
the sum of contributions of single cells (synergy)
~\cite{Vaadia95,DeCharms96,Riehle97,Singer97, Maynard99}, or
negative (redundancy)~\cite{Gawne93,Zohary94,Shadlen98}.
Thus, specific examples can  be found of correlations that limit the 
fidelity of signal transmission, and others that carry additional 
information. Another view is that usually correlations do not make 
much of a difference in either direction~\cite{Golledge96, Amit97, Rolls98}
and that their presence can be regarded  as a kind of random 
noise. In this paper, we show that even when correlations are of a completely
random nature they may contribute very substantially, and to some extent
predictably, to information transmission.
 
To discuss this point we must first  quantify the amount of information 
contained in the neural response. Information theory~\cite{Shannon48}
provides one framework for describing mathematically the process of 
information transmission, and it has been applied successfully to the 
analysis of neuronal recordings~\cite{Rolls98, Eckhorn74, Optican87, Rieke96}.
Consider a stimulus taken from a finite discrete set 
$\cal{S}$ with $S$ elements, each stimulus $s$ 
occurring with probability $P(s)$. The 
probability of response ${\bf r}$ (the ensemble activity, imagined 
 as  the firing rate vector) is $P({\bf r})$, and the joint probability
distribution is $P(s,{\bf r})$. The mutual information between 
the set of stimuli $\cal{S}$
and the response is
\eq
I(t) = \sum_{s \in \cal{S}}\sum_{{\bf r}}P(s,{\bf r}) 
\log_2 \frac{P(s,{\bf r})}{P(s) P({\bf r})}
\label{shannon}
\en
where $t$ is the length of the time window for recording the response 
${\bf r}$.

We study here the contribution of correlations to such mutual information. 
In the $t\to 0$ limit, the mutual 
information can be broken down into a firing rates and correlations
components, as shown by Panzeri {\it et al.}~\cite{Pan+99} and summarized  
in the next section. The correlation-dependent part can be further 
expanded by considering  ``small'' correlation coefficients 
(see Section~\ref{Large}). In this (additional) limit approximation
the effects of correlations can be analyzed and it will be seen that even 
if they are random they give large contributions to the total information. 
The number of second-order (pairwise) correlation terms in the information 
expansion in fact grows as $C^2$, where $C$ is the number of cells, 
while contributions that depend only on individual cell firing rates 
 of course grow linearly with $C$.
As a result, as shown by Panzeri {\it et al.}~\cite{NCip},
the time window to which the expansion is applicable
shrinks as the number of cells increases, and conversely the overall 
effect of correlation grows.
We complement this derivation by analysing (see Section~\ref{rats}) 
the response of cells in the rat somatosensory barrel cortex 
during the response  to deflections of the vibrassae.
Conclusions about the general applicability of correlation 
measures to information transmission are drawn in the last section.

\section{The short time expansion}
In the limit $t\to 0$, 
following Ref.~\cite{Pan+99},  
the information carried by the population response can be expanded
in a Taylor series
\eq I(t) = t\; I_t + {t^2\over 2}\; I_{tt} + \dots
\label{Taylor}
\en
There is no zero order term, as no information can be 
transmitted in a null time-window.
Higher order terms (see Ref.~\cite{Laura00}) are not considered here, but 
they could be included in a straightforward extension of this approach.

Assuming that the conditional probability of
a spike being emitted by  cell $i$, $n_i=1$, given 
that cell $j$ has fired scales proportionally to $t$, i.e.:
\begin{equation}
P(n_i(s)=1|n_j=1;s) \equiv \overline{r}_{i}(s) t (1+\gamma_{ij}(s))
\label{pn}
\end{equation}
(where the $\gamma_{ij}(s)$ coefficient quantifies correlations) 
the expansion~(\ref{Taylor}) becomes an expansion in the total number of spikes
emitted by an assembly of neurons (see Ref.~\cite{Pan+99} for details).
Briefly,
the procedure is the following:
the  expression for $P(n)$~(\ref{pn}) is inserted
 in the Shannon formula for 
information, Eq.(\ref{shannon}), whose  logarithm is then expanded 
in a power series. All  terms, with the same power of $t$, are grouped 
and compared to Eq.(\ref{Taylor}), to extract first and
second order derivatives.

The first time derivative (i.e. the information rate)
depends only on the firing rates averaged over trials with the same stimulus, 
denoted as $\overline{r}_{i}(s)$

\eqa 
I_t &=& \sum_{i=1}^C \left< \overline{r}_{i}(s) \log_2
{\overline{r}_{i}(s)\over
\left<\overline{r}_{i}(s')\right>_{s'}}~\right>_{s}
\label{eq:1stderiv}
\nonumber
\ena
i.e. the sum of the information rate of each single cell~\cite{Ska+92,Bia+91}.

The second derivative breaks into three components 
\[
I_{tt}= I_{tt}^{(1)} + I_{tt}^{(2)} + I_{tt}^{(3)}
\]

These terms depend on two different kinds of correlations, usually 
termed `signal' and `noise' correlations~\cite{Gawne93}.
`Noise' correlations are pairwise correlations in the response 
variability, i.e. a measure of the tendency of both of the cells to fire more
(or less) during the same trial, compared to their average response 
over many trials with the same stimulus.
For short time windows this is a measure of the synchronization 
of the cells.
We introduce the `scaled cross-correlation density'~\cite{Aer+89},
i.e. the amount of trial by trial concurrent  firing
between different cells, compared to that 
expected in the uncorrelated case
\[
\gamma_{ij}(s) = {\overline{{r}_i(s) {r}_j(s)}\over
\overline{r}_i(s)\overline{r}_j(s) } -1.
\]
This coefficient can vary from $-1$ to $+\infty$; negative 
values indicate anticorrelation, whereas positive $\gamma_{ij}$ values
indicate correlation.
For $i=j$, the `scaled autocorrelation coefficient' $\gamma_{ii}$ gives
the probability of observing a spike emission, given that the same 
cell has already fired in the same time window; i.e. 
\[
\gamma_{ii}(s) = {\overline{{r}^{2}}_i(s) - \overline{r}_i(s)\over
\overline{r}_i^2(s)} -1
\]
The relationship with alternative cross-correlation coefficients, like 
the Pearson correlation, is discussed in Ref.~\cite{Pan+99}. 

`Signal' correlations measure the tendency of pairs of cells to
respond more (or less) to the same stimuli in the stimulus set.
As in the previous case we introduce the signal
cross-correlation coefficient $\nu_{ij}$,
\begin{equation}
\nu_{ij} = {<\overline{r}_i(s) \overline{r}_j(s) >_s \over
<\overline{r}_i(s)>_s <\overline{r}_j(s)>_s} -1
\label{nudef}
\end{equation}
and, similarly, we define the autocorrelation coefficient $\nu_{ii}$.

The first term of $I_{tt}$ again depends only on the mean rates:
\[
I_{tt}^{(1)}  =  
{1\over \ln 2} \sum_{i=1}^C  \sum_{j =i}^C 
\left<\overline {r}_{i}(s)\right>_{s} \left<\overline
{r}_{j}(s)\right>_{s}  \biggl[  
\nu_{ij} + (1 + \nu_{ij})\ln ({1\over 1+\nu_{ij}} ) \biggr]
\]

The second term  is non-zero only when correlations are  present both 
in the noise 
(even if they are stimulus-independent) and in the signal
\[
I_{tt}^{(2)}= {1\over \ln 2}
\sum_{i=1}^C  \sum_{j =i}^C \biggl[ \left< \overline{r}_{i}(s)
\overline{r}_{j}(s)  
\gamma_{ij}(s) \right>_s 
\biggr] \ln ({1\over 1+\nu_{ij}}).  
\]

The third term contributes only if correlations are stimulus-dependent
{
\eqa
I_{tt}^{(3)} &=& {1\over \ln 2} 
\sum_{i=1}^C  \sum_{j =i}^C \left< \overline{r}_{i}(s)
\overline{r}_{j}(s) (1 + \gamma_{ij}(s)) 
\ln \left[{ \left<\overline{r}_{i}(s')
\overline{r}_{j}(s')\right>_{s'} (1+ \gamma_{ij}(s))\over  
\left<\overline{r}_{i}(s') 
\overline{r}_{j}(s')(1+ \gamma_{ij}(s'))\right>_{s'} } \right] \right>_s .
\nonumber 
\ena
}

The sum $I_t t + \frac{1}{2} I_{tt}^{(1)} t^2$ depends only on 
average firing rates, $\overline{r}_{i}(s)$,
(rate only contribution) and its first term is always 
greater than or equal to zero, while $I_{tt}^{(1)}$ is always less
than or equal to zero.

In the presence of correlations, i.e. non zero $\gamma_{ij}$
and $\nu_{ij}$, more information may be available when observing
simultaneously the responses of many cells, than when observing them
separately: {\em synergy}. For two cells, it can happen due to  positive 
correlations in the variability, if the mean rates to different stimuli are
anticorrelated, or vice-versa. If the signs of signal and noise
correlations are the same, the result is always {\em redundancy}. 
Quantitatively, the impact of correlations is minimal when the 
mean responses are only weakly correlated across the stimulus set.

The time range of validity of the expansion~(\ref{Taylor}) is limited
by the requirement that second order terms be small with respect to
first order ones, and successive orders be negligible.
Since at order $n$ there are $C^n$ terms with $C$ cells,
the applicability of the short time limit contracts 
for larger populations. 

\section{Large number of cells}
\label{Large}
 
Let us investigate  the 
role of correlations in the transmission of information by  large 
populations of cells. 
For a few cells, all cases of synergy or
redundancy are possible if the correlations are properly engineered,
in simulations, or if, in experiments, the appropriate special case is recorded. The outcome of the information analysis simply
reflects the peculiarity of each case. With large populations, one 
may hope to have a better grasp of generic, or typical, cases, more
indicative of conditions prevailing at the level of, say, a given
cortical module, such as a column.

 Consider a `null' hypothesis model of a large population:
purely {\em random} correlations; i.e. correlations that were not designed 
to play any special role  in the system  being analyzed.

In this null hypothesis, signal correlations $\nu_{ij}$
can be thought of as arising from a random walk with $S$ steps (the number of 
stimuli). Such a random walk of positive and negative steps 
 typically spans a range of size $\sqrt{S}$. The $\nu_{ij}$
 have zero average, 
while the squares $\nu_{ij}^2$ still differ from zero on average, 
since they are positive. Noise correlations may 
be thought to arise from stimulus-independent terms, $\gamma_{ij}$
(which may well be large), and from stimulus-dependent contributions, which
we denote $\delta 
\gamma_{ij}(s)$ and  which might be expected to get smaller when more trials per
stimulus are available, and whose squares again would be expected to span
the range of a random walk (whose steps are now the different trials).
The effect of such null hypothesis correlations
on information transmission can be gauged by further expanding $I_{tt}$ in the 
small parameters $\nu_{ij}$ and $\delta \gamma_{ij} (s)$, i.e. 
assuming $|\nu_{ij}| << 1$ and $|\delta \gamma_{ij} (s)| << 1$.

Consider, first, the expansion of $I_{tt}^{(1)}$, 
that does not depend on $\gamma_{ij} (s)$. Expanding in powers of
$\nu_{ij}$ and neglecting terms of order $3$ or higher, we easily 
get:
\eqa
I_{tt}^{(1)} & = & 
-{1\over \ln 2} \sum_{i=1}^C  \sum_{j =i}^C 
\left<\overline {r}_{i}(s)\right>_{s} \left<\overline
{r}_{j}(s)\right>_{s}  \nu_{ij}^2
\label{I_1f}  
\ena

The second contribution of $I_{tt}$, up to the second order in $\nu_{ij}$,
is 
\eqa
I_{tt}^{(2)}&=& -{1\over \ln 2}
\sum_{i=1}^C  \sum_{j =i}^C \biggl[
(1+\gamma_{ij})\left<\overline {r}_{i}(s)\right>_{s} \left<\overline
{r}_{j}(s)\right>_{s} \nu_{ij}^2 +\\
&+&(1+\gamma_{ij})\left<\overline {r}_{i}(s)\right>_{s} \left<\overline
{r}_{j}(s)\right>_{s} \nu_{ij} + \left< \overline{r}_{i}(s)
\overline{r}_{j}(s)  
\delta \gamma_{ij}(s) \right>_s 
\nu_{ij}\biggr] 
\label{I_2f} 
\ena

The third contribution, $I_{tt}^{(3)}$ is more complicated, as 
an expansion in   $\delta \gamma_{ij}(s)$ is required as well.
Expanding the logarithm in these small parameters
up to second order we get:
\eqa
I_{tt}^{(3)} &=& {1\over {\ln 2 }} 
\sum_{i=1}^C  \sum_{j =i}^C {1\over {1+ \gamma_{ij}}}
\biggl[\left< \overline{r}_{i}(s)
 \overline{r}_{j}(s) \delta\gamma_{ij}(s)^2 \right>_s
-\frac{\left< \overline{r}_{i}(s)
\overline{r}_{j}(s) \delta\gamma_{ij}(s) \right>^2_s}{
\left< \overline{r}_{i}(s)
\overline{r}_{j}(s)\right>_s} \biggr]
\label{I_3f} 
\ena

Introducing the average on stimuli weighted on the 
product of the normalized firing rates, $\overline{r}_{i}(s)
 \overline{r}_{j}(s)/\left< \overline{r}_{i}(s)
\overline{r}_{j}(s)\right>_s$, that is  
\[
\left<\delta\gamma_{ij}(s)\right>_{\{i,j\},s} = 
\frac{1}{S} \sum_{i=1}^S \frac{\overline{r}_{i}(s)
 \overline{r}_{j}(s)}{
\left< \overline{r}_{i}(s)
\overline{r}_{j}(s)\right>_s}\delta\gamma_{ij}(s)
\]
we obtain, from Eq.~(\ref{I_3f}), 
\[
I_{tt}^{(3)}={1\over {\ln 2}} 
\sum_{i=1}^C  \sum_{j =i}^C 
\frac{(\nu_{ij}+1)\left< \overline{r}_{i}(s)\right>_s
\left< \overline{r}_{j}(s) \right>_s}{1+ \gamma_{ij}}
\biggl[\left<\delta\gamma_{ij}(s)^2\right>_{\{i,j\},s}
-\left<\delta\gamma_{ij}(s)\right>_{\{i,j\},s}^2 \biggr]
\]
that is a non-negative quantity, i.e. a synergetic contribution to information.
In case of random ``noise'' correlations, with zero weighted average
over the set of stimuli, 
i.e. $\left<\delta\gamma_{ij}(s)\right>_{\{i,j\},s}=0$, 
this equation can be re-written, 
\begin{equation}
I_{tt}^{(3)} = {1\over {\ln 2}} 
\sum_{i=1}^C  \sum_{j =i}^C 
\frac{\nu_{ij}+1}{1+ \gamma_{ij}}\left< \overline{r}_{i}(s)\right>_s
\left< \overline{r}_{j}(s) \right>_s
 \left<\delta\gamma_{ij}(s)^2\right>_{\{i,j\},s} 
\equiv 
\frac{C(C+1)}{2 \ln2} \left<\delta\gamma^2 \right>_C
\label{I_3fm} 
\end{equation}
where we have introduced, to simplify the notation, 
\[
\left<\delta\gamma^2 \right>_C \equiv \frac{1}{C(C+1)/2}
\sum_{i=1}^C  \sum_{j =i}^C 
\frac{\nu_{ij}+1}{1+ \gamma_{ij}}\left< \overline{r}_{i}(s)\right>_s
\left< \overline{r}_{j}(s) \right>_s
\left<\delta \gamma_{ij}(s)^2\right>_{\{i,j\},s}
\]

Assuming purely random signal correlations $\nu_{ij}$ with zero average, 
we get, summing eqs.~(\ref{I_1f}) and~(\ref{I_2f}),
\begin{equation}
I_{tt}^{(1)}+I_{tt}^{(2)}= -{2\over \ln 2}
\sum_{i=1}^C  \sum_{j =i}^C 
(1+\frac{\gamma_{ij}}{2})\left<\overline {r}_{i}(s)\right>_{s} \left<\overline
{r}_{j}(s)\right>_{s} \nu_{ij}^2 \equiv -
\frac{C(C+1)}{\ln 2}\left< \nu^2\right>_C 
\label{red}
\end{equation}
where we have introduced 
(in a similar way as for $\delta\gamma$; these
two definitions 
coincide for $\nu_{ij}$ and $\gamma_{ij}\rightarrow 0$): 
\[
\left<\nu^2\right>_C =\frac{1}{C(C+1)/2}
\sum_{i=1}^C  \sum_{j =i}^C 
(1+\frac{\gamma_{ij}}{2})\left<\overline {r}_{i}(s)\right>_{s} \left<\overline
{r}_{j}(s)\right>_{s} \nu_{ij}^2
\]
This contribution (Eq.~\ref{red}) to information 
is always negative ({\it redundancy}).

Thus the leading contributions of the new Taylor expansion are 
of two types, both coming as $C(C+1)/2$ terms proportional to 
$\left<\overline{r}_{i}(s)\right>_s
\left< \overline{r}_{j}(s) \right>_s$. The first one, Eq.~(\ref{red}), is a 
redundancy term proportional to $\left<\nu^2\right>$;  
the second one, Eq.~(\ref{I_3fm}), is a synergy term
roughly proportional to $\left<\delta \gamma^2\right>$. 

These leading contributions to $I_{tt}$ can be compared 
to first order contributions to the original Taylor expansion in $t$ (i.e.,
to the $C$ terms in $I_t$) in different time ranges. For times $t\approx 
{\rm ISI}/C$, that is $t\left<\overline{r}\right>\approx 1/C$, first order 
terms sum up to be of order one bit, while second order terms are smaller 
(to the extent that $\left<\nu^2\right>_C$ and $\left<\delta \gamma^2\right>_C$
are taken to be small). This occurs
however over a time range that becomes shorter as more cells are
considered, and the total information conveyed by the population remains
of order 1 bit. 

For times $t \simeq {\rm ISI}$, i.e. $t\left<\overline{r}\right>\approx 1$, 
first order terms are of order $C$, while second order ones are of order 
$C^2\left<\nu^2\right>_C$ (with a minus sign, signifying redundancy) and 
$C^2 \left<\delta \gamma^2\right>_C$ (with a plus sign, signifying synergy) 
respectively. If $\left<\nu^2\right>_C$ and
$\left<\delta \gamma^2\right>_C$ 
are not sufficiently small to counteract the additional
$C$ factor, these ``random'' redundancy and synergy contributions will
be substantial. Moreover, over the same time ranges  leading contributions
to $I_{ttt}$ and to the next terms in the Taylor expansion in time may
be expected to be substantial. The expansion already will be of limited use 
by the time  most cells have  fired just a single spike.

If this bleak conclusion comes from a model with small and random
correlations, what is the time range of applicability of the expansion
when several real cells are recorded simultaneously?

\section{Measuring correlations in rat barrel cortex}
\label{rats}
We  have analyzed many sets of data recorded from rat cortex. 
Part of the primary somatosensory cortex of the rat (the barrel 
cortex) is organized in a grid of columns, with each column anatomically and
functionally
associated with one homologous whisker (vibrissa)
on the controlateral side: the column's neurons respond maximally, on average,
 to the 
deflection of this ``principal'' whisker. In our experiments,
in a urethane-anesthetized rat, one whisker  was stimulated 
at 1 Hz, and each deflection lasted for
100 $ms$. The latency (time delay between stimulus onset and the 
evoked response) in this fast sensory system is usually around $5-10 ms$. 
We present here 
the complete analysis of a single typical dataset.
The physiological methods are described in Ref.~\cite{Mathew}.

\begin{figure}[h]
\centerline{
\hbox{
\psfig{figure=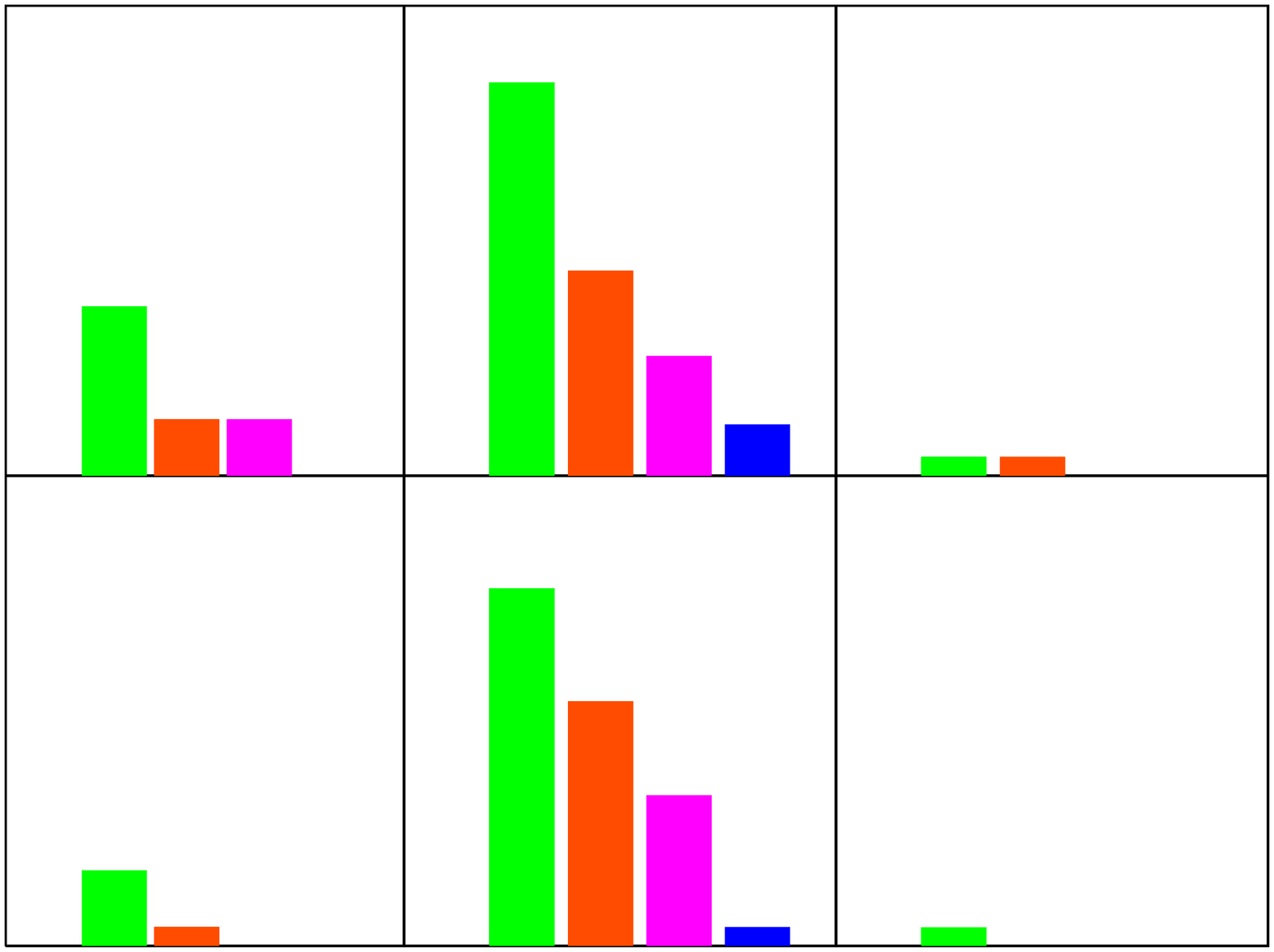,width=5.5cm,angle=270}
\psfig{figure=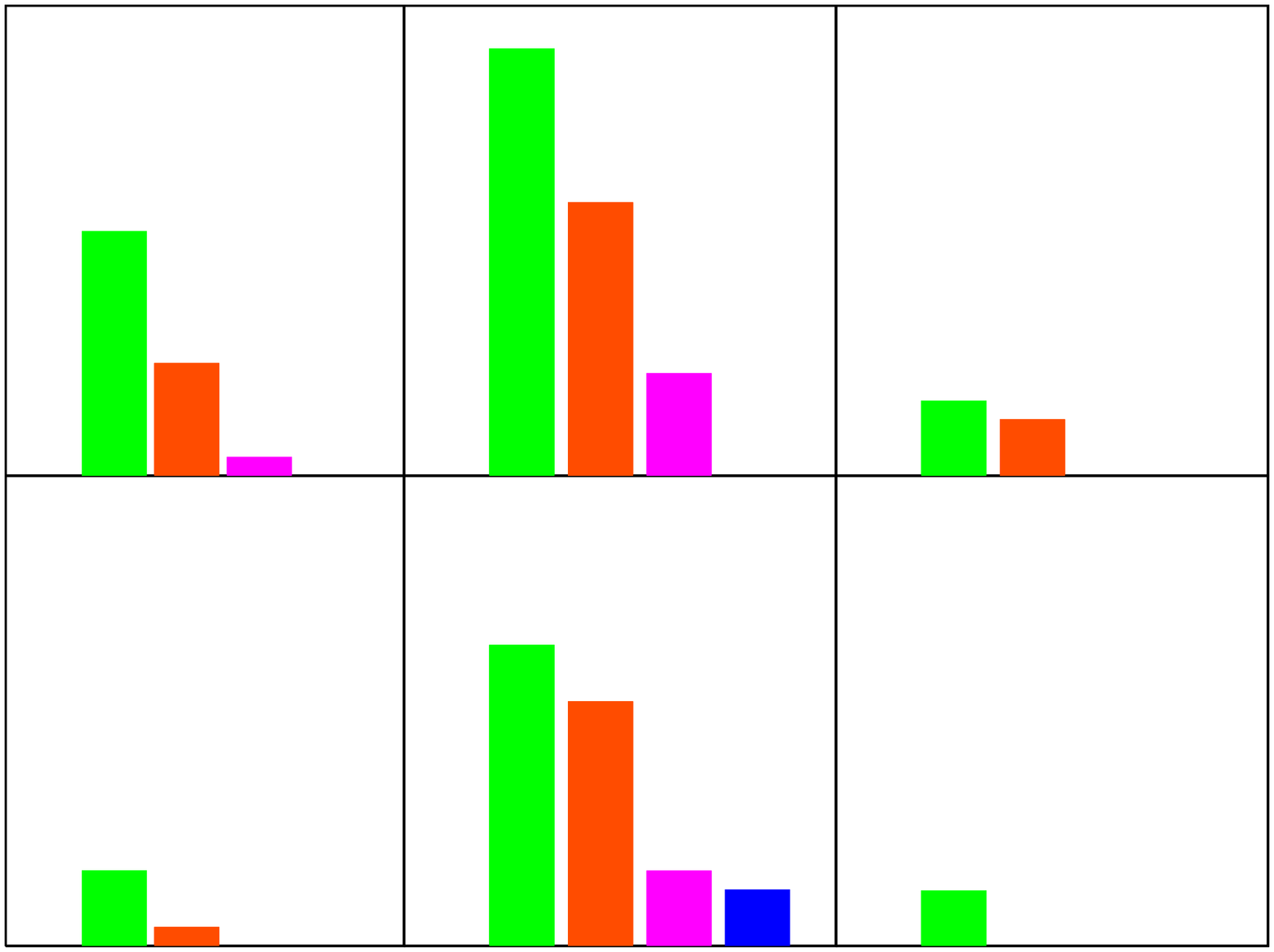,width=5.5cm,angle=270}
\psfig{figure=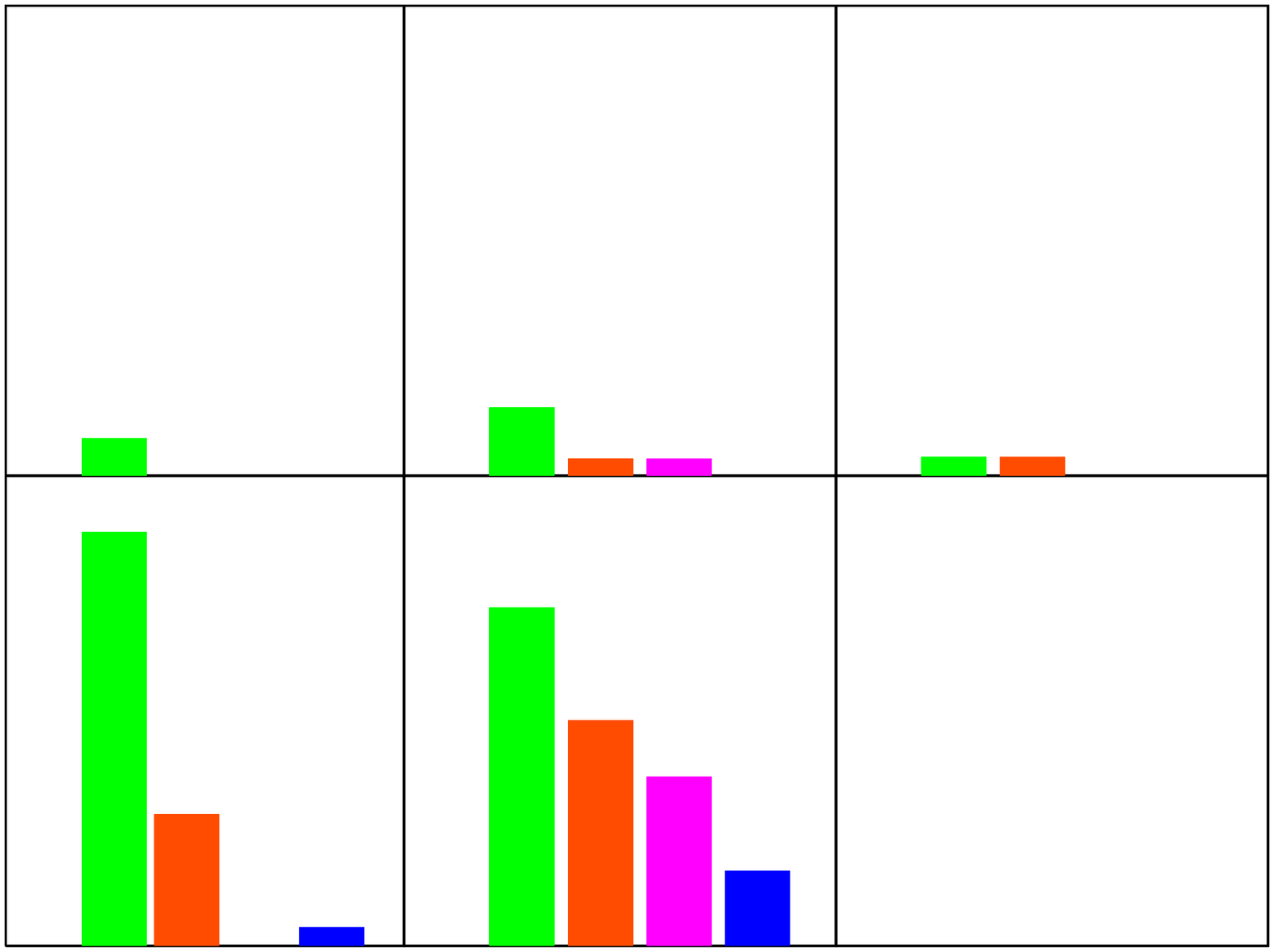,width=5.5cm,angle=270}
}}
\centerline{
\hbox{
\psfig{figure=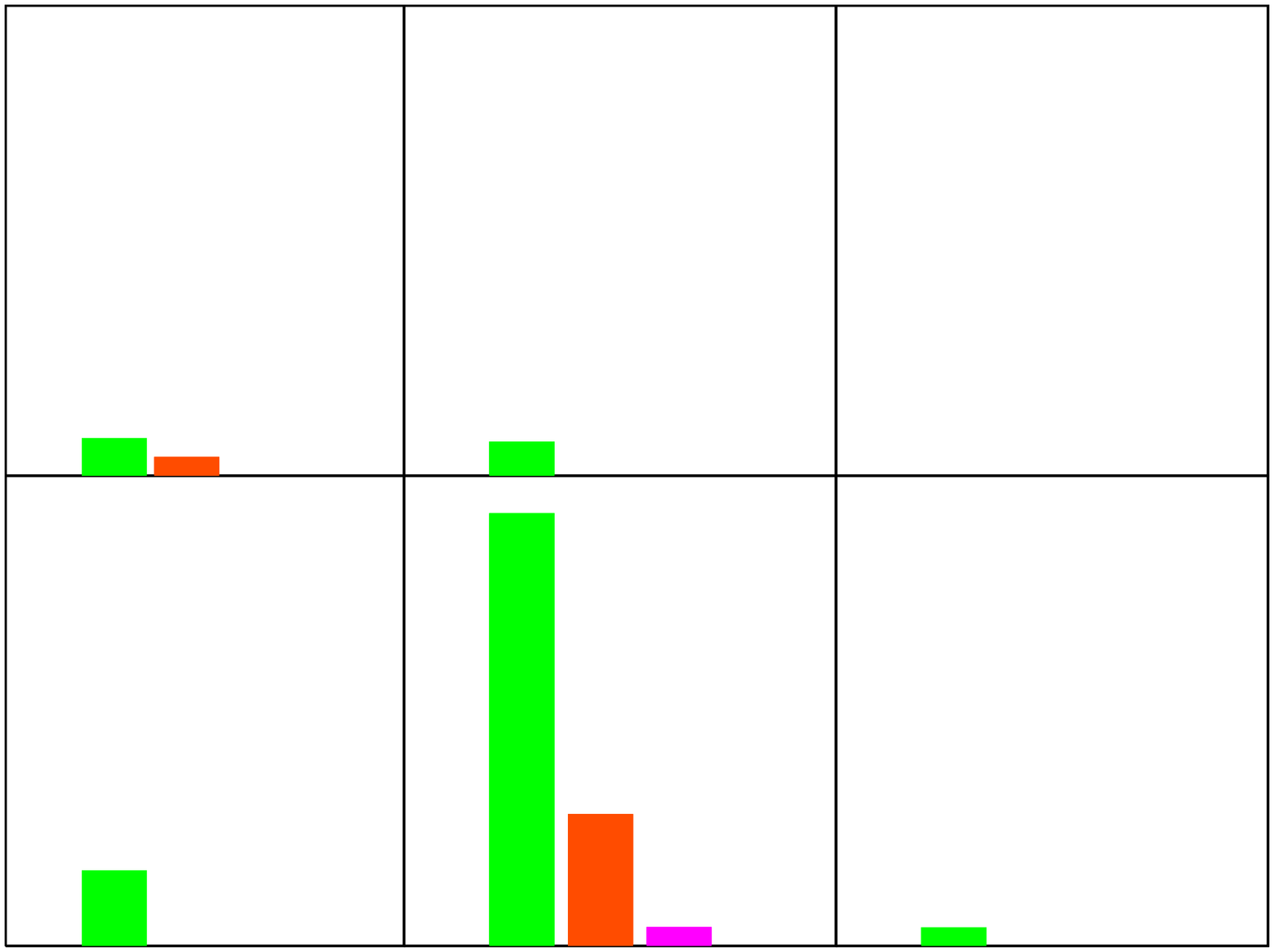,width=5.5cm,angle=270}
\psfig{figure=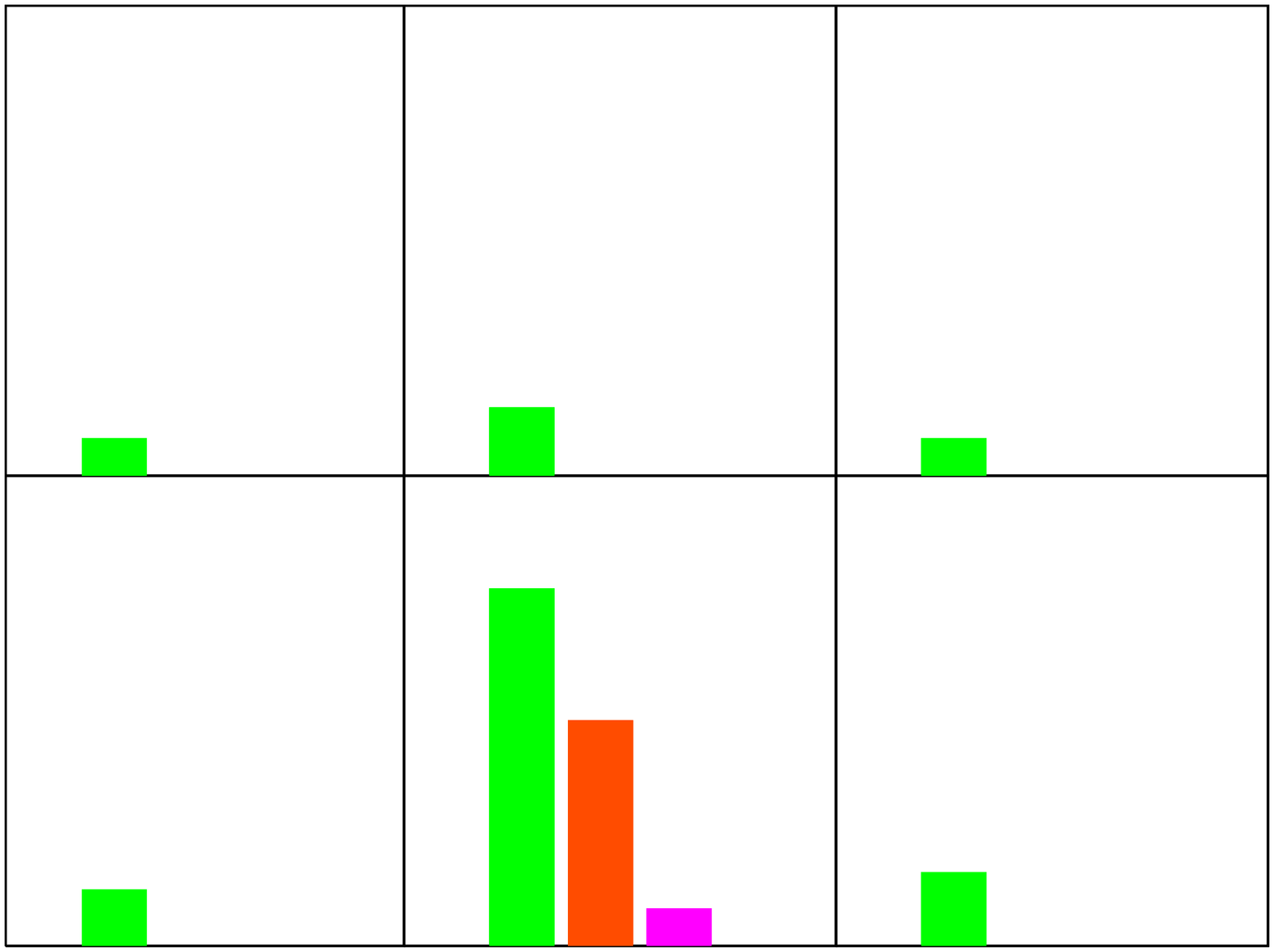,width=5.5cm,angle=270}
\psfig{figure=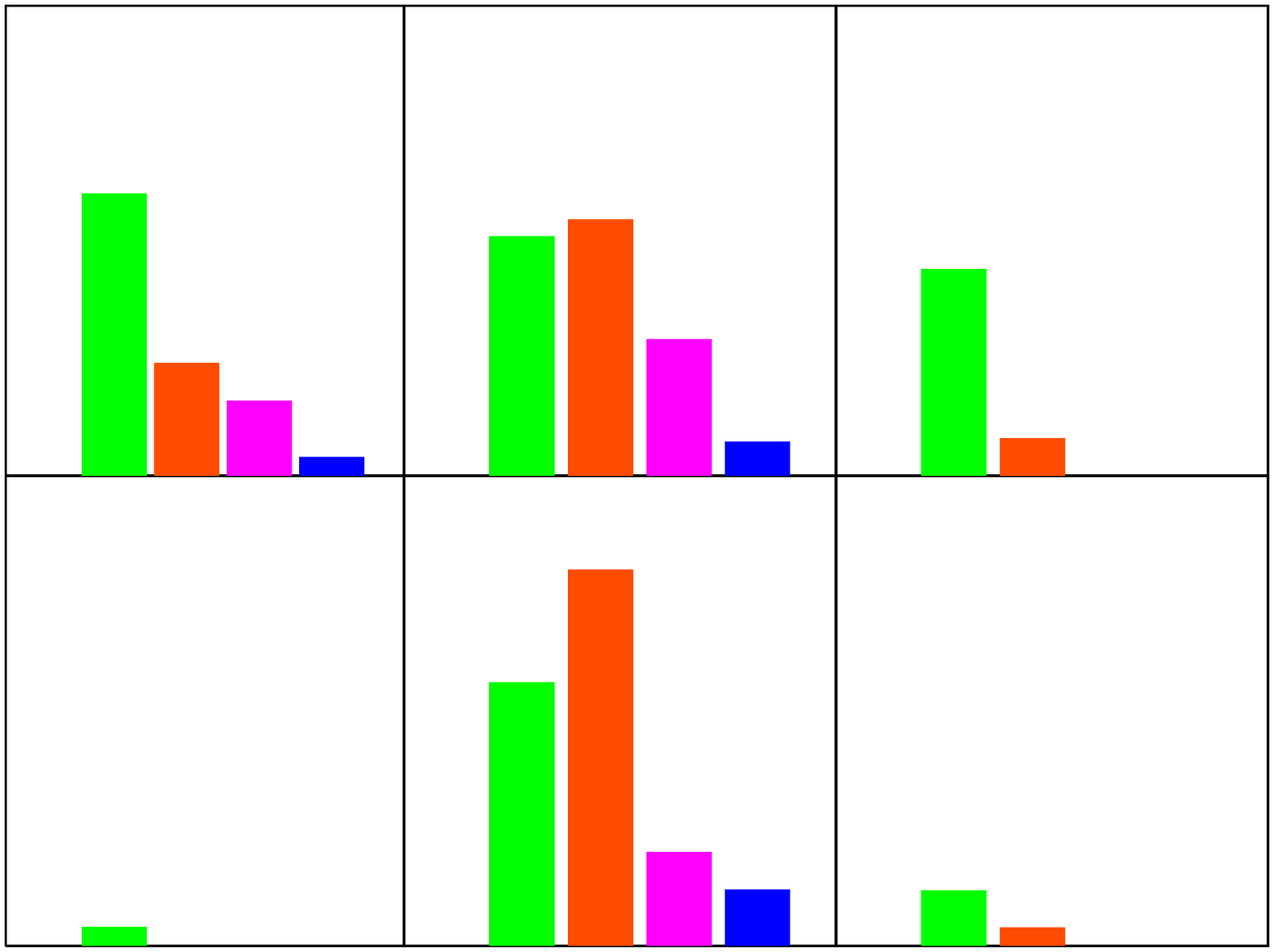,width=5.5cm,angle=270}
}}
\centerline{
\hbox{
\psfig{figure=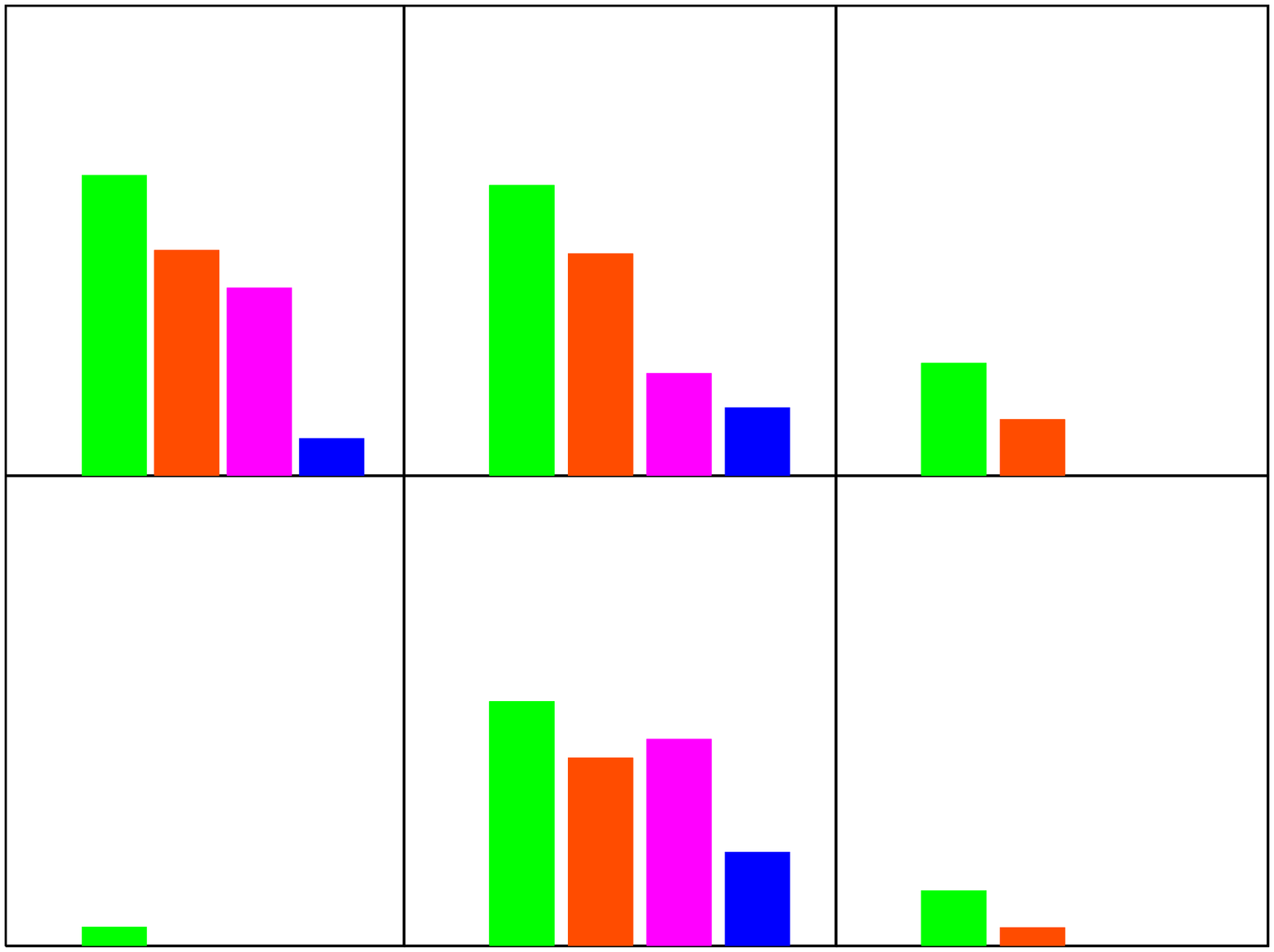,width=5.5cm,angle=270}
\psfig{figure=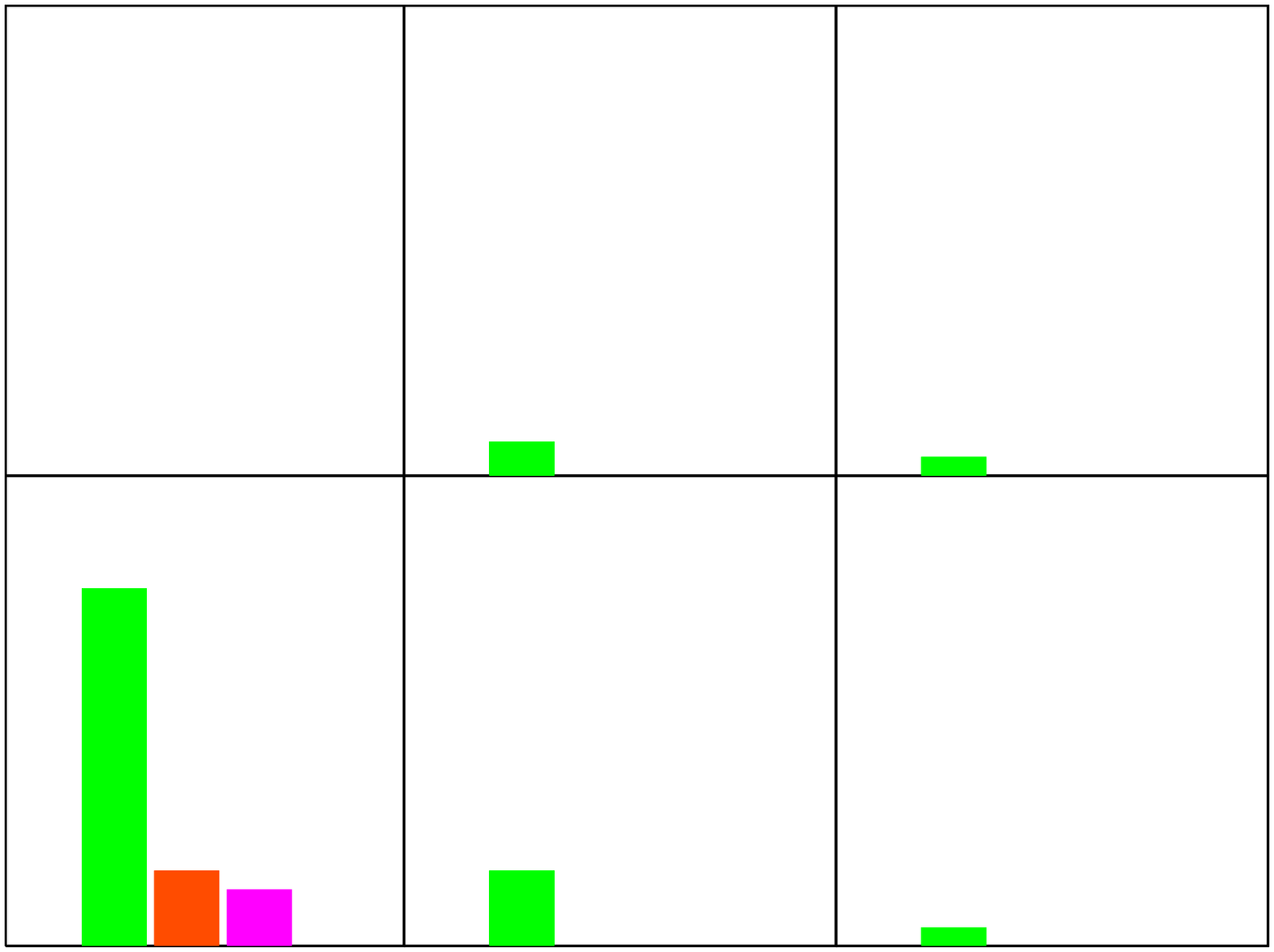,width=5.5cm,angle=270}
\psfig{figure=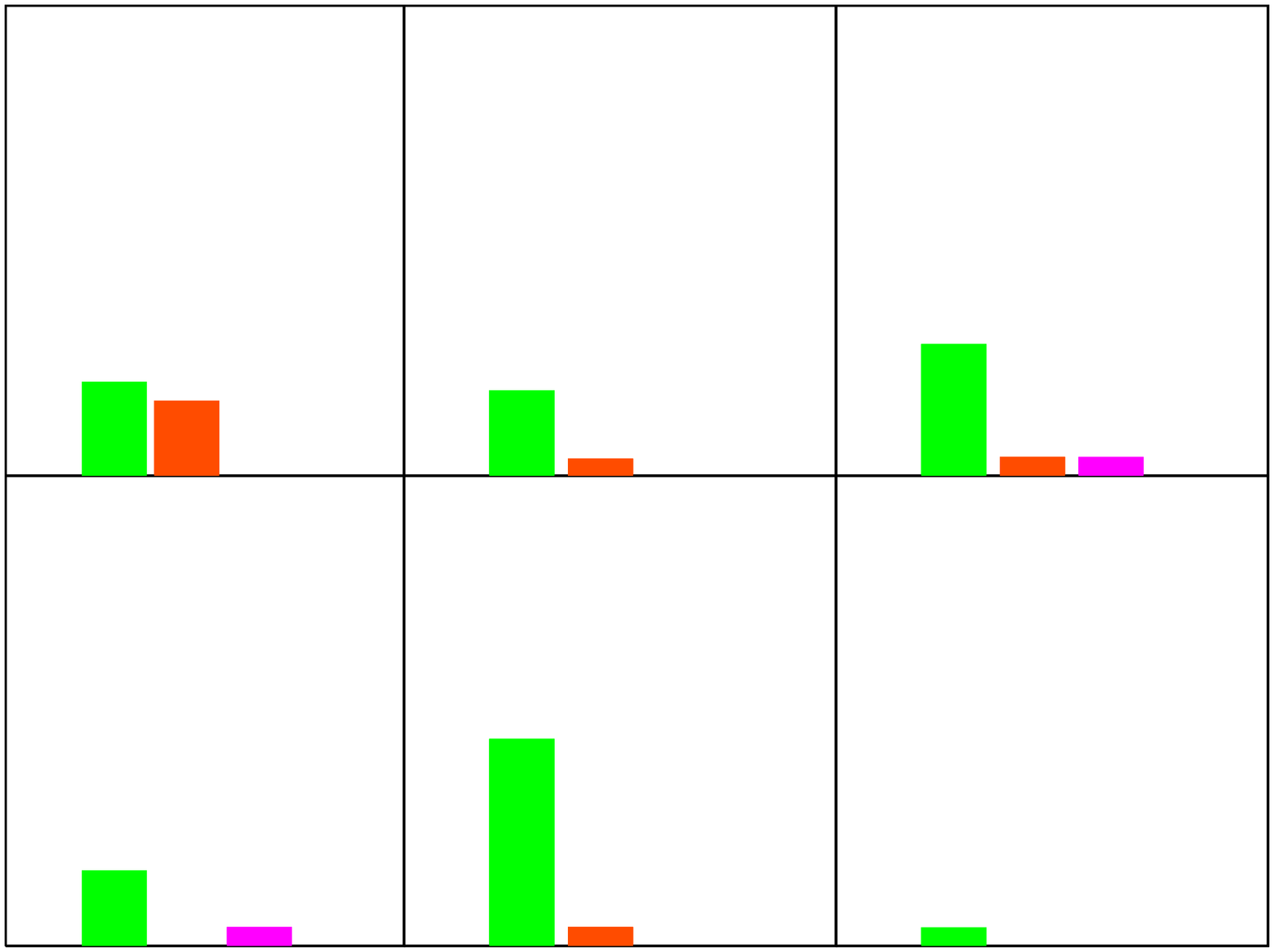,width=5.5cm,angle=270}
}}   
\caption{Firing rate distributions (probability of observing 
a given number of spikes emitted in the time window of $40ms$),
for 9 cells and 6 different stimuli (that is, $6$ whisker sites). The bars, from left to 
right, represent the probability to have  ($0$ not shown) 
$1$, $2$, $3$, or more than $3$ (black bar) spikes during a single stimulus 
presentation.
Maximum $y$-axis set to $0.5$.
}
\label{rate}
\end{figure}
For each stimulus site there were   $50$ trials and in our analysis 
we have considered up to $6$ stimulus sites,
(i.e. different whiskers) with 
12  cells recorded simultaneously.
In Fig.~\ref{rate} we report the firing distributions of 9 of the 12 
cells for each of the 6 stimuli. 
One can immediately note that several cells are most strongly activated 
by a single whisker, while responding more weakly or not at all to
the others. 
Other cells have less sharply tuned receptive fields. 
A mixture of sharply tuned and more broadly tuned receptive fields is
characteristic of a given population of barrel cortex neurons.
We have computed the distribution of $\nu_{ij}$ and $\delta 
\gamma_{ij}(s)$ for different time windows. 

\begin{figure}[h]
\centerline{
\psfig{figure=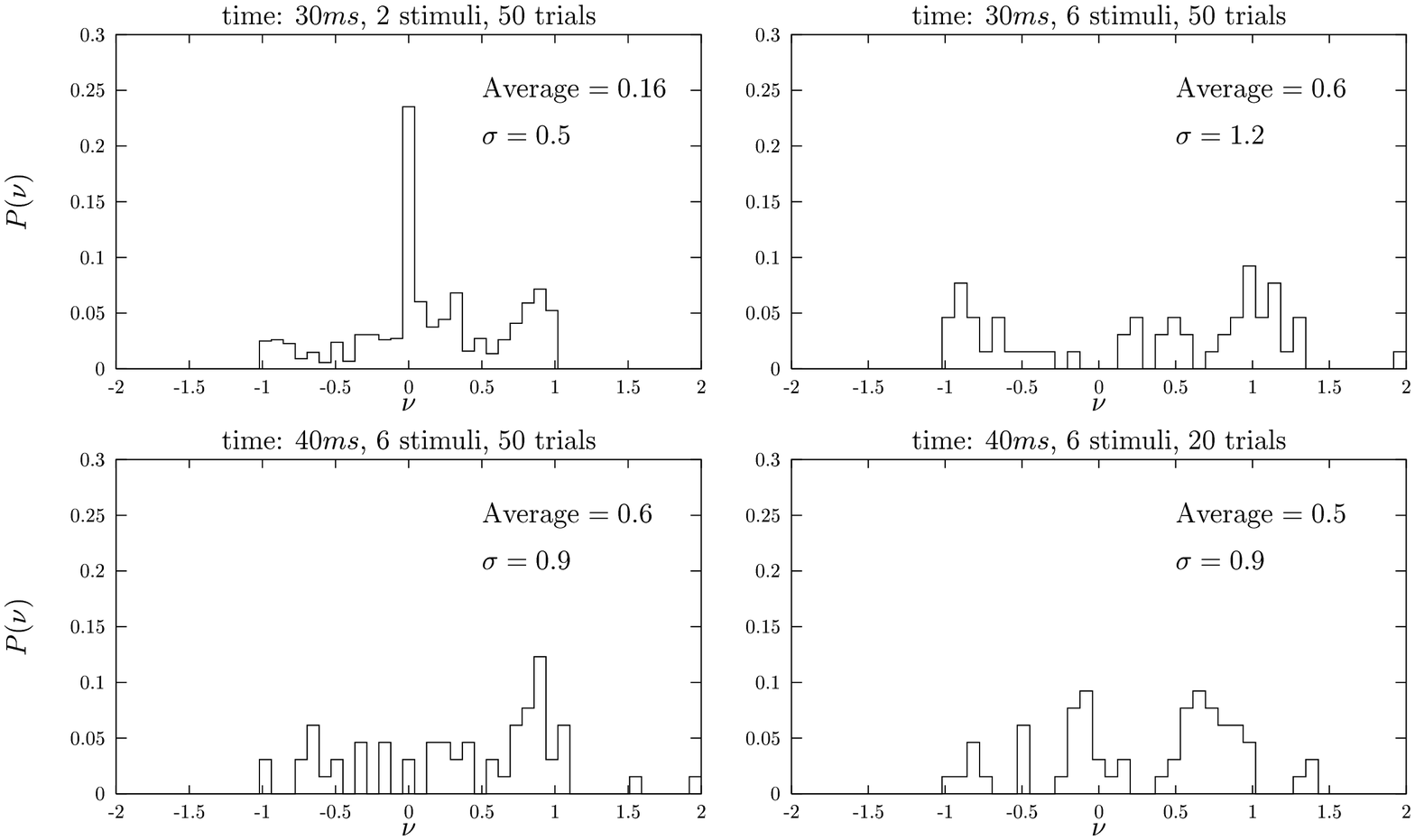,width=16cm} 
}
\caption{The distribution $P(\nu)$, considering the $\nu_{ij}$ for each
cell pair $i$ and $j$. Computed after $30ms$ with 2  stimuli, $50$ trials 
per stimulus (averaged over all possible pairs
from among $6$ stimuli, top left),  with all the $6$ stimuli (top right), 
with $6$ stimuli and after $40ms$ (bottom left) and with  $6$ stimuli 
and after $40ms$ but with only $20$ trials per stimulus (bottom right).}
\label{nu}
\end{figure}
In Fig.~\ref{nu} we have plotted the distribution of all $\nu_{ij}$.
In the first figure (Fig.~\ref{nu}, top, left) we have considered 2 stimuli,
taken from the set of $6$ stimuli, and averaged over all the possible pairs.
In the following figure (Fig.~\ref{nu}, top, right), where we take all the $6$
stimuli, the distribution is broader ($\sigma=1.2$ {\em vs.} $\sigma=0.5$ in
the previous case), and the maximum value of $\nu$ is $2.8$
({\em vs.} $1.0$). This larger spread of $\nu$ values can be explained
by the fact that most cells have a greater response for one stimulus, and
weaker for the other. 
This can be seen by considering a limit case: two cells $i$ and $j$ fire 
just to a single stimulus $s'$, i.e. $\overline{r}_{i}(s') \neq 0$
$\overline{r}_{j}(s') \neq 0$ and
$\overline{r}_{j}(s)= \overline{r}_{i}(s)=0$ for $s \neq s'$.
From Eq.~(\ref{nudef}), we have 
\[
\nu_{ij} = {\frac{1}{S}\sum_s\overline{r}_i(s) \overline{r}_j(s)  \over
\frac{1}{S}\sum_s \overline{r}_i(s) \frac{1}{S}\sum_s\overline{r}_j(s)} -1
={\frac{1}{S}\overline{r}_i(s') \overline{r}_j(s')  \over
\frac{1}{S}\overline{r}_i(s') \frac{1}{S}\overline{r}_j(s')} -1
= S-1
\]
As the total number of stimuli 
$S$ increases, $\nu$ values of the order of $S$ appear, and broaden the distribution.
The distribution does not change qualitatively when the time window lengthens,
(Fig.~\ref{nu}, bottom, left) at least from $30ms$ to $40ms$, except
for a somewhat narrower width with the longer time-window. For very 
short time windows ($\leq 20ms$) we observe instead a peak at $\nu=0$ and $\nu=-1$ 
due to the prevalence of cases of zero spikes: when the mean rates of at least
one of the two cells are zero to all stimuli $\nu=0$, and when the stimuli, to
which each gives a non-zero response, are mismatched, then $\nu=-1$.
In the last panel (Fig.~\ref{nu}, bottom, right) we have taken a more limited 
sample ($20$ trials), which in this case does not significantly change  the 
moments of the distribution.
 
\begin{figure}[h]
\centerline{
\psfig{figure=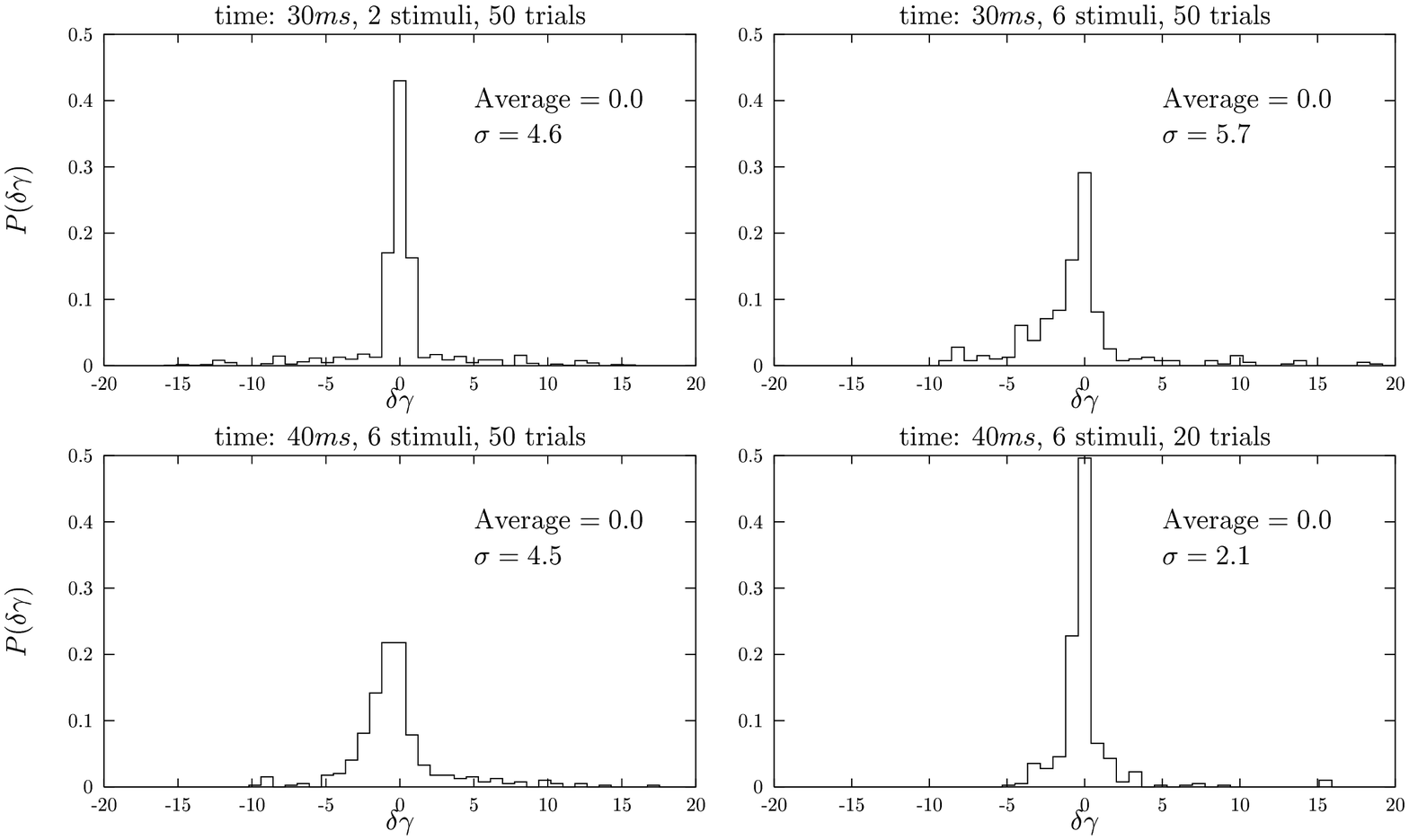,width=16cm} 	
}
\caption{The distribution $P(\delta\gamma)$, considering all 
$\delta\gamma_{ij}$ for each cell pair $i$ and $j$ and for each stimulus $s$. 
Computed after $30ms$ with 2 stimuli, $50$ trials per stimulus (averaged over 
all possible pairs of $6$ stimuli, top left), with all the $6$ stimuli (top 
right), with $6$ stimuli and after $40ms$ (bottom left) and with with $6$ 
stimuli and after $40ms$ but with only $20$ trials per stimulus (bottom right).}
\label{deltagamma}
\end{figure}
The distribution of $\delta \gamma$ is illustrated in Fig.~\ref{deltagamma}.
This distribution has $0$ average by definition; we can observe
that the spread becomes larger when increasing the number of stimuli 
from $2$ (Fig.~\ref{deltagamma}, top left) to $6$ 
(Fig.~\ref{deltagamma}, top right). This derives from having
rates $r_{i}(s)$ that differ from zero 
and only for one or a few stimuli. 
In this case increasing the number of stimuli the fluctuations in the 
distribution of $\gamma_{ij}$ (and hence of $\delta \gamma_{ij}(s)$) become 
larger, broadening the distribution.
For longer time windows (Fig.~\ref{deltagamma}, bottom, left), there are more 
spikes and a better sampling of the rates, so the spread of the distribution
decreases ($\sigma=4.5$ for $40ms$ {\em vs.} $\sigma=5.7$ for $30ms$). 
The effect of finite sampling ($20$ trials)  illustrated in the 
last plot (Fig.~\ref{deltagamma}, bottom, left), is  now 
a substantial reduction in width.

\begin{figure}	
\centerline{
	\psfig{figure=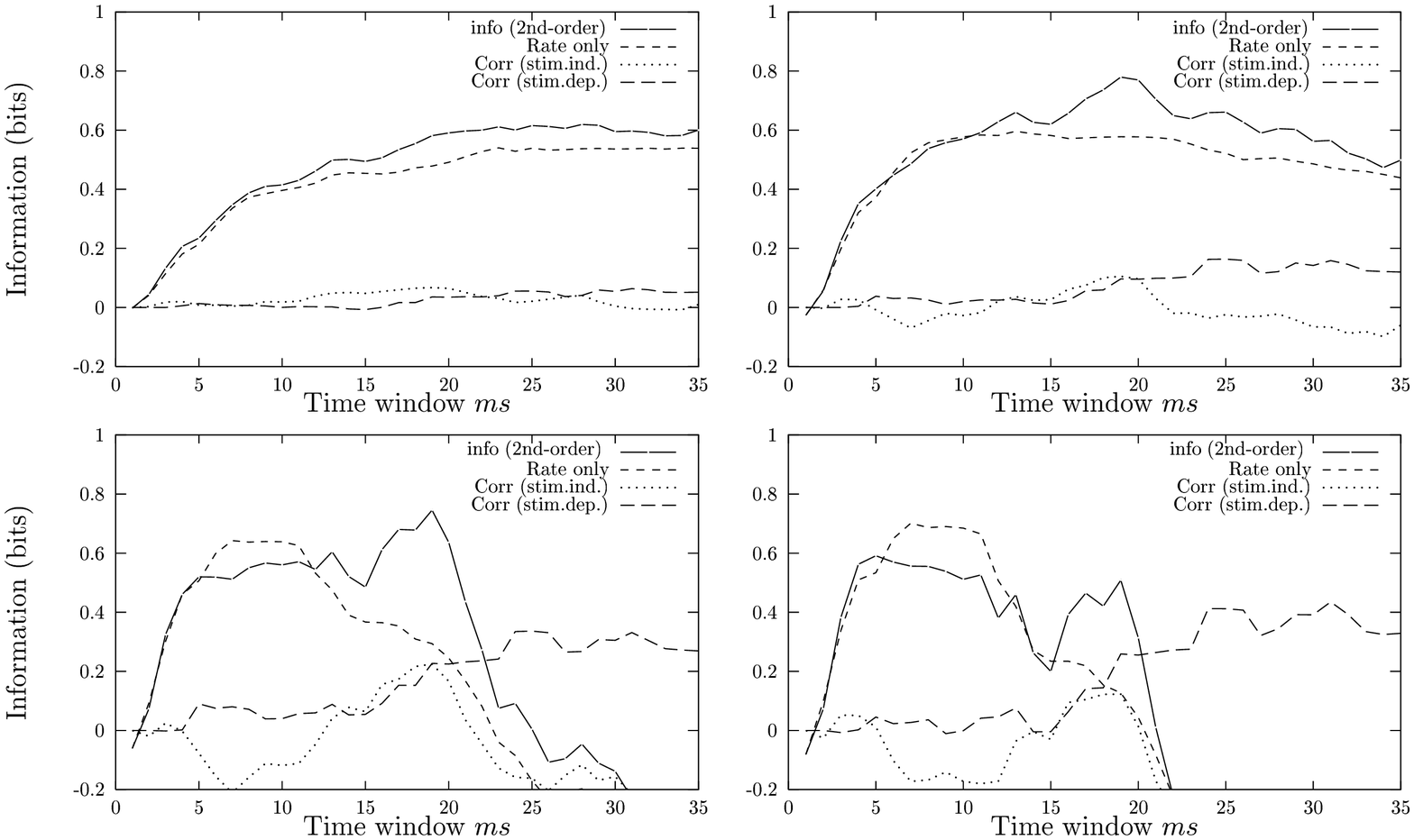,width=18cm} 
} 
\caption{The short time limit expansion breaks down sooner when
the larger  population is considered. Cells in rat somatosensory
barrel cortex for  $2$ stimulus sites. 
Components of the transmitted information (see text for details) with $3$  (top, left),
$6$ (top, right), $9$ (bottom, left) and $12$ cells (bottom, right).
The initial slope (i.e. $I_t$) is roughly proportional to the number of cells.
The effects of the second order terms, quadratic in $t$, are visible over the
brief times
between the linear regime and the break-down of the expansion.
Information is estimated taking into account finite sampling
effects~\protect{\cite{Pan+96}}. Time window starts $5ms$ after the stimulus
onset.}
\label{fig:rats}
\end{figure} 
In Fig.~\ref{fig:rats} we have plotted, for the same experiment, 
the values of the information and of single terms of the second order 
expansion discussed above. 
The full curve represents the  information $I(t)$ up to the second order,
i.e. $I(t) = t\; I_t + {t^2\over 2}\; I_{tt}$
The short dashed curve is the sum $I_t t + \frac{1}{2} I_{tt}^{(1)} t^2$,  
i.e. the {\it rate only} contribution, as it depends only on 
average firing rates, $\overline{r}_{i}(s)$. The dotted line 
represents $\frac{1}{2} I_{tt}^{(2)} t^2$, 
i.e. the contribution of correlations to 
information even if they are {\it stimulus independent}.
The last second-order term, $\frac{1}{2} I_{tt}^{(3)} t^2$, is non-zero
only if the correlations are {\it stimulus dependent}.
We expect 
$I_t$ to grow linearly with the number of cells, and in fact
 the slope of the total information (full curve in Fig.~\ref{fig:rats})
increases linearly, at least for the short time interval before 
the second derivative starts to bend it down. 
As mentioned above, the number of second order terms grows as $C^2$, causing 
the range of validity of the expansion up to second order 
to decrease with the number of cells, as evident in Fig.~\ref{fig:rats}.
Note that, in general, as the number of cells increases one would
 expect an increase in 
the  information  they conveyed, which is {\it not} what one observes 
in Fig.~\ref{fig:rats}, except in the brief initial linear regime.
This is an indication of the failure of the second order expansion, which
for $12$ cells appears to break down after little more than $5ms$ from the 
response onset.

\section{Conclusions}

The contribution of pairwise correlations to the total information 
conveyed by the activity of an ensemble of cells can 
be quantified by introducing a Taylor expansion of the information
transmitted over cumulative time intervals, and calculating its terms up to 
second order. The range of validity of the expansion depends on the 
overall magnitude of second order terms with respect to first order ones.
We have shown, by considering a model with `small' random correlations
in a large ensembles, 
that for times $t \simeq {\rm ISI}$ (inter-spike interval),
 the expansion would already 
begin to break down. The overall contribution of first order terms is in fact 
of order $C$, while second order ones are of order $-C^2\left<\nu^2\right>_C$ 
(redundancy) and $C^2 \left<\delta \gamma^2\right>_C$ (synergy).
These `random' redundancy and synergy contributions will normally be 
substantial, unless a specific mechanism minimizes the values of  
$\left<\nu^2\right>_C$  and $\left<\delta \gamma^2\right>_C$
well below order $1/C$.

Further, data from the somatosensory cortex of the rat indicate that
the assumption of `small' correlations may be far too optimistic in the real
brain situation; the expansion may then break down even sooner, 
although one should consider that the rat somatosensory cortex is a ``fast''
 system, with short-latency responses and high instantaneous firing rates.
 
Our data show (see Fig.~\ref{fig:rats}) that the range of validity
of the second-order expansion decreases approximatively as $1/C$.
The length of the time interval over which the expansion is valid is
roughly $10-15ms$ 
for $9$ or $12$ cells, in agreement with Panzeri and 
Schultz~\cite{NCip}.
They have found, analyzing a large amount of cells recorded from the somatosensory 
barrel cortex of an adult rat, that for single cells 
the expansion works well up to $100ms$.
In its range of validity this expansion constitutes an efficient tool
for measuring information even in the presence of limited sampling of data,
when a direct approach using the full Shannon formula, 
Eq.(\ref{shannon}), turns out to be impossible~\cite{Pan+99}.
When its limits are respected, the expansion can be used to address fundamental
questions, such as extra information in timing and the relevance of correlations
contribution.

It is important  that if second order terms are
comparable in size to first order terms, all successive orders in the
expansion are also likely to be relevant. The breakdown of the expansion 
is then not merely a failure of the mathematical formalism, but an
indication that this particular attempt to quantify, in absolute terms, the information
conveyed by a large ensemble is intrinsically ill-posed in that time range.
There might be other expansions, or other ways to measure mutual 
information e.g. the reconstruction method~\cite{Bia}, that
lead to better results.

A pessimistic conclusion is then that the expansion should be applied only
to very brief times, of the order of $t\approx {\rm ISI}/C$. In this
range the information rates of different cells add up independently,
even if cells are highly correlated, but the total information conveyed,
no matter how large the ensemble, remains of order 1 bit.
 
A more optimistic interpretation stresses the importance
of considering information decoding along with information encoding.
In this vein, not all pairwise correlations are taken into account
on the same footing, and similarly not all correlations to higher orders;
rather, appropriate models of neuronal decoding prescribe which variables can 
affect the activity of neurons downstream, and it is only a limited
number of such variables that are included as corrections into the
evaluation of the information conveyed by the ensembles. This embodies
the assumption that real neurons may not be influenced by the information (and
the synergy and redundancy) encoded in a multitude of variables that cannot 
be decoded. In an ideal world, it would be preferable to 
characterize the quantity of information present in population activity and to 
assume that the target neurons can conserve all such information. In real life, such an
assumption does not seem to be justified, and considerable further work is now needed 
to explore different models of neuronal decoding, and their implementation
in estimating information, in order to make full use of the potential
offered by the availability of large scale multiple single-unit recording
techniques.

\section*{Acknowledgments}

This work was supported in part by HFSP grant RG0110/1998-B, and is a
follow-up to the analyses of the short time limit, in which Stefano
Panzeri has played a leading role. We are grateful to him, and to Simon
Schultz, also for the information extraction software, and to Misha Lebedev 
and Rasmus Petersen for their help with the experimental data.
The physiology experiments were supported by NIH grant NS32647 to M.E.D.

\end{document}